\documentclass[
 reprint,
nofootinbib,
 amsmath,amssymb,
 aps,
]{revtex4-2}

\usepackage{aasmacros}
\usepackage[utf8]{inputenc}
\usepackage{graphicx}
\usepackage{dcolumn}
\usepackage{bm}
\usepackage{newtxtext,newtxmath}
\usepackage{subcaption}
\usepackage[T1]{fontenc}
\usepackage{makecell}
\usepackage{animate}
\usepackage{mathtools}
\usepackage{url}
\usepackage[hidelinks,colorlinks]{hyperref}
\usepackage{xcolor}
\hypersetup{
    colorlinks=true,
    linkcolor={blue},
    citecolor={blue},
    urlcolor={blue}
}

\begin{document}

\preprint{APS/123-QED}

\title{Convergence Tests of Self-Interacting Dark Matter Simulations}

\author{Charlie Mace$^{1,2}$}\email{E-mail: mace.103@osu.edu}
\author{Zhichao Carton Zeng$^{1,2}$}%
\author{Annika H. G. Peter$^{1,2,3,4}$}%
\author{Xiaolong Du$^{5,6}$}
\author{Shengqi Yang$^{6}$}
\author{Andrew Benson$^{6}$}
\author{Mark Vogelsberger$^{7}$}
\affiliation{$^{1}$Department of Physics, The Ohio State University, 191 W. Woodruff Ave., Columbus OH 43210, USA \\
$^{2}$Center for Cosmology and Astroparticle Physics, The Ohio State University, 191 W. Woodruff Ave., Columbus OH 43210, USA\\
$^{3}$Department of Astronomy, The Ohio State University, 140 W. 18th Ave., Columbus OH 43210, USA\\
$^{4}$School of Natural Sciences, Institute for Advanced Study, 1 Einstein Drive, Princeton, NJ 08540\\
$^{5}$ Department of Physics and Astronomy, University of California, Los Angeles, CA 90095, USA\\
$^{6}$ Carnegie Observatories, 813 Santa Barbara Street, Pasadena CA 91101, USA\\
$^{7}$ Department of Physics, Massachusetts Institute of Technology, 77 Massachusetts Avenue, Cambridge, MA 02139, USA\\
}

\date{\today}

\begin{abstract}
Self-interacting dark matter (SIDM) theory predicts that dark matter halos experience core-collapse, a process where the halo’s inner region rapidly increases in density and decreases in size. The N-body simulations used to study this process can suffer from numerical errors when simulation parameters are selected incorrectly. Optimal choices for simulation parameters are well studied for cold dark matter (CDM), but are not deeply understood when self-interactions are included. In order to perform reliable N-body simulations and model core-collapse accurately we must understand the potential numerical errors, how to diagnose them, and what parameter selections must be made to reduce them. We use the \texttt{Arepo} N-body code to perform convergence tests of core-collapsing SIDM halos across a range of halo concentrations and SIDM cross-sections, and quantify potential numerical issues related to mass resolution, timestep size, and gravitational softening length. Our tests discover that halos with fewer than $10^5$ simulation particles, a resolution typically not met by subhalos in N-body simulations, suffer from significant discreteness noise that leads to variation and extreme outliers in the collapse rate. At our lowest resolution of $N=10^4$ particles, this collapse time variation can reach as high as 20\%. At this low resolution we also find a bias in collapse times and a small number of extreme outliers. Additionally, we find that simulations which run far beyond the age of the Universe, which have been used to calibrate SIDM gravothermal fluid models in previous work, have a sensitivity to the timestep size that is not present in shorter simulations or simulations using only CDM. Our work shows that choices of simulation parameters that yield converged results for some halo masses and SIDM models do not necessarily yield convergence for others.

\end{abstract}

\maketitle

\section{Introduction}

Self-interacting dark matter (SIDM) is a class of dark matter models that include interactions between dark matter particles \cite{spergel00,battaglieri17,tulin18,BUCKLEY18,adhikari22}. These models, which have theoretical motivation from particle physics \cite{feng2009,tulin2013a,tulin2013b,boddy2014,cline2014,FYCR2016}, may provide solutions to several unresolved contradictions between cold dark matter (CDM) and observational data, such as the diversity problem (CDM; \cite{KuziodeNaray_2014,oman15,errani18,read19,Relatores_2019,santos20,Hayashi_2020,Li_2020}, SIDM; \cite{kaplinghat16,kamada17,sameie20,correa21,carton2022,correa22,Nadler_2023,dYang_2023}) and the cuspy halo problem (CDM; \cite{flores94,moore94,NFW97,moore99,weinberg15}, SIDM; \cite{Kaplinghat_2020,Bhattacharyya_2022}). There are many models for these self-interactions, some with elastic scattering and some with inelastic scattering. The inelastic models produce their own distinct phenomenology \cite{chua20,oneil23,roy23}, but for this work we will focus solely on the elastic scattering case. In elastic models, the additional heat transfer channel provided by self-interactions causes the SIDM halos to undergo a time evolution not experienced by CDM halos, where an initially cuspy halo thermalizes and forms an isothermal, low-density, constant density core \cite{Colín_2002}. The core formation process of an SIDM halo is followed by a slow increase in the halo's core density, a phase called core-collapse. During this phase the core transfers energy to the outer regions of the halo, causing particles to fall into the core. This collapse, also called the gravothermal catastrophe, is a runaway process that eventually causes a rapid density increase in the center of the halo \cite{Kochanek:2000pi,balberg2002}. The phenomenon was first discovered in the study of globular clusters, where gravitational two-body scatterings of stars provide the heat transfer instead of dark matter self-interaction \cite{lyndenbell1980}, but the gravothermal catastrophe model has found success in describing dark matter halos as well.

The study of SIDM core-collapse depends on reliable N-body simulations. One use of simulations is to investigate how self-interactions influence the formation of substructure  \cite{rocha2013,kahlhoefer2019,correa22,carton2022,carton2023}, but N-body simulations are also essential to the study of the simplest case of an isolated core-collapsing halo. The density profile of such a halo has been shown to follow a self-similar time evolution in the regime where the mean free path between collisions is much longer than the gravitational scale height (the long mean free path regime) \cite{balberg2002,koda2011,essig_SIDM,Nishikawa:2019lsc,Outmezguine:2022bhq,shengqi2023,gadnasr2023}. These density profile models, which are found from numerical solutions of gravothermal fluid equations, rely on a constant factor that must be calibrated using N-body simulations \cite{balberg2002,koda2011,essig_SIDM}. This factor is needed as the conductivity used in the long mean free path gravothermal fluid solution is based on the thermal conductivity of a gas, which does not include the gravitational dynamics present in a dark matter halo. While we do not have an SIDM conductivity derived from first principles for the long mean free path regime, this approximation of using the thermal conductivity of a gas has been shown to be accurate up to a constant factor on the heat transfer rate \cite{balberg2002,koda2011,essig_SIDM}. This heat transfer constant, referred to as $\beta$ in this work but sometimes represented by $C$ in the literature, must be calibrated to N-body simulations.

The use of N-body simulations to understand SIDM core-collapse is complicated by numerical errors in these simulations. There has been extensive study of numerical parameter selection in CDM N-body simulations \cite{moore1998,power2003,vdb2018_2,vdB2018}, and recent work has extended this analysis to SIDM \cite{meskhidze22,zhong2023,palubski2024numerical,fischer2024numerical}, but how these choices fare when self-interactions are included is still not fully understood. SIDM simulations may not converge for the same parameter choices as CDM, and spurious evolution that is well understood in CDM may look entirely different when combined with the core-collapse process. One possible difference, for example, is in timestep selection. N-body simulations are typically based on symplectic N-body algorithms, which do not produce diverging numerical errors \cite{Leimkuhler_Reich_2005}. The typical prescription for symplectic integration requires the simulation to operate on a global, time-invariant timestep. CDM simulations typically employ adaptive and variable time-stepping schemes, which permit each simulation particle to have an independent, time-dependent timestep size. This violates the assumptions made when formulating a symplectic integrator, which means these simulations may suffer from secular energy and momentum error accumulation \cite{quinn97,GADGET4}. For CDM this problem is manageable, provided the system can be approximated as collisionless \cite{gadget2}. SIDM simulations, however, are not collisionless, and may require more care in timestep assignment.

In this work we conduct a detailed study of numerical error and parameter selection in SIDM N-body simulations using the N-body code \texttt{Arepo} \cite{arepo2010} with a previously tested SIDM module \cite{vogelsberger2012,vogelsberger13,vogelsberger14}. Our study investigates how three numerical parameters (softening length, timestep size, and particle number) affect the core-collapse process in constant cross-section SIDM, and what values for these parameters provide converged results. This investigation is performed for two different halo concentrations that are each tested over a two different collapse timescales, creating four scenarios. These four cases provide insight into how concentration and core-collapse time must be considered when selecting numerical parameters. Additional factors that may affect the numerical errors, such as alternate timestepping schemes, will be studied in future work.

Two other investigations of SIDM convergence in N-body simulations were published shortly after the first version of this study was completed~\cite{palubski2024numerical,fischer2024numerical}. Both our work and \cite{palubski2024numerical} investigate how SIDM simulations respond to changes in halo concentration, mass resolution, and timestep size, finding similar general results. Our work focuses on a wide breadth of convergence testing, including cross-section variation and fixed force softening length variation. For an analysis of other factors, such as different scattering implementations and adaptive force softening, we direct readers to \cite{palubski2024numerical}. The investigation in \cite{fischer2024numerical} includes the study of additional choices, such as adaptive versus fixed timesteps, and the opening criterion in tree-based gravitational force evaluations.

In Section \ref{sec:method} we outline our simulation strategy, and describe the timestepping and gravothermal modeling in more detail. Section \ref{sec:noiseTests} presents tests of the initial condition realization noise, and our primary convergence testing results are in Section \ref{sec:CT}. Our findings and recommendations for parameter selection are summarized in Section \ref{sec:conclusion}.

\section{Method}

\label{sec:method}

In this section we provide details on how we run and analyze our simulations in \texttt{Arepo}. First we describe the gravothermal fluid model that we use in our parameter selection and simulation analysis, and define important variables we use to describe our simulation sets. Next we give details on our timestepping algorithm and associated numerical parameters, and describe our convergence testing parameter grid. Lastly, we present the definitions and rationale behind the central density (which serves as a tracer of the core-collapse process) and collapse time measurements used in core-collapse analysis.

\subsection{Gravothermal Fluid Modelling}

In this work we often compare our simulations to predictions made by gravothermal fluid models. Gravothermal collapse can be modeled by treating the dark matter as a fluid, and assuming that fluid evolves slowly enough that it remains in hydrostatic equilibrium. The fluid equations for the system can then be solved numerically, providing a density profile prediction as a function of time \cite{balberg2002,koda2011,Pollack:2014rja,essig_SIDM,Nishikawa:2019lsc,Outmezguine:2022bhq,shengqi2023}. This prediction depends on the strength of self-interactions in the halo, quantified locally by the Knudsen number $Kn\equiv\lambda/H$. Here we define $\lambda$ as the mean free path of self-interactions and $H$ is the local Jeans length $H\equiv 2\pi v/\sqrt{4\pi G \rho}$, where $v$ and $\rho$ are the local velocity dispersion and mass density respectively. The Jeans length is used here since it is the approximate size of a self-gravitating system \cite{BinneyTremane_GD}, telling us the radial distance a given particle will typically traverse in the halo. We can approximate the strength of self-interactions globally by the dimensionless scattering cross-section $\hat{\sigma}\equiv(\sigma/m)\rho_s r_s$, where $\rho_s$ and $r_s$ are the characteristic density and radius of the initial Navarro-Frenk-White (NFW) profile \cite{NFW96} characteristic of CDM halos.

When $\hat{\sigma}\ll 1$ for a halo, we say it is in the long mean free path (lmfp) limit. In this regime, the distance between scattering events is much larger than the local Jeans length for particles in the halo. This means these particles complete many orbits between each scattering event. When $\hat{\sigma}\gg 1$, the halo is in the short mean free path (smfp) limit. In this regime a particle's motion is dominated by frequent scattering events, causing a more fluid-like behavior. While $\hat{\sigma}$ is a global parameter that has no dependence on position in the halo, $Kn$ will actually depend on the local density, which will vary throughout the halo. A typical SIDM halo will have an outer region that is in the lmfp regime, and a central core that may be in either regime depending on the particular parameters and current stage of core-collapse, which means a full gravothermal collapse analysis must consider the intersection of the two regimes \cite{gadnasr2023}. The $\hat{\sigma}$ variable is an approximate metric for whether the halo is generally in the lmfp or smfp regime during the core formation phase.

The calibration parameter $\beta$ is introduced in the thermal conductivity in the lmfp regime. The heat transfer coefficient, also called the conductivity, at a given position in the dark matter halo can be written as:

\begin{equation}
    \kappa_{\mathrm{lmfp}}=\frac{3\beta}{2}\frac{nH k_B}{t_r}
\end{equation}
where $n$ is the local number density, $H$ is the local Jeans length, $k_B$ is the Boltzmann constant, and $t_r$ is the local relaxation time \cite{lyndenbell1980,essig_SIDM}. The SIDM relaxation time is defined as the mean time between collisions for a particle, and can be calculated from the local mass density $\rho$, local velocity dispersion $v$, and cross-section per unit mass $\sigma$ as $t_r\equiv(\sqrt{16/\pi}\rho\sigma v)^{-1}$\cite{balberg2002}. The original derivation of this conductivity for globular clusters in \cite{lyndenbell1980} was based on heat conduction in a gas, and introduced the $\beta$ parameter (referred to as $C$ in that work) to account for any discrepancy in the approximation caused by the gravitationally bound orbits of stars.

Solving the fluid equations in the lmfp regime yields the following model for the halo central density:

\begin{equation}
\begin{split}
    \log(\rho_{c}/\rho_s) &= A(\log(\beta\hat{\sigma}\hat{t}) - (E+3))^2 + C \\
    &+ \frac{D}{(E+0.0001-\log(\beta\hat{\sigma}\hat{t})^{0.02}}
\end{split}
\label{eqn:rhocpred}
\end{equation}
where $A=0.05771$, $C=-21.64$, and $21.11$, and $E=2.238$ are dimensionless parameters fit to the numerical result, and $\hat{t}$ is a dimensionless time parameter defined as $\hat{t}\equiv t\sqrt{4\pi G \rho_s}$. This model includes $\beta$ as a free parameter, which must be calibrated to N-body simulations as it has not been derived from first principles. Further details of this core-collapse model can be found in Appendix A of \cite{shengqi2023}.

Throughout this work we compare N-body central density evolutions to the gravothermal prediction Equation \ref{eqn:rhocpred}, which relies on the calibration parameter $\beta$. We calibrate $\beta$ individually for each set of physical parameters using our simulation data to compare our simulations to the fluid model. Since Equation \ref{eqn:rhocpred} is based on the lmfp regime there is no guarantee it will describe the full range of dark matter halos included in this study, but after $\beta$ calibration we find the agreement close enough for our purposes. Calibrating to the lmfp solution when some of our simulations could have a substantial smfp contribution to the heat transfer means our $\beta$ values may not necessarily represent the ``true'' value, and are instead only effective $\beta$ values for this particular method of calibration. Readers should be cautious when comparing our values to other values in the literature, as the effects of calibrating in this intermediate regime are not yet well understood. In future work, we will perform a systematic calibration of $\beta$ across a range of halo parameters and SIDM parameter space, and test where this model fails.

\subsection{Timestepping}

There are two timestep assignments used in our \texttt{Arepo} simulations: the acceleration-based timestep and the SIDM timestep. The values for these timesteps are evaluated for each particle whenever a new step is taken, and that particle uses the smaller value of the two.

The acceleration-based timestep:
\begin{equation}
    \delta\mathrm{t}_a = \sqrt{\frac{2\eta r_{\mathrm{soft}}}{|\boldsymbol{a}|}}.
    \label{eqn:dta}
\end{equation}
is based on the gravitational acceleration $\boldsymbol{a}$ of the particle, and the particle's softening length $r_{\mathrm{soft}}$. This assignment decreases the timestep for rapidly accelerating particles to reduce numerical error. The dimensionless parameter $\eta$ is selected as needed to scale the timestep size, with a default value of 0.025 in the \texttt{GADGET-4} documentation\footnote{Available at \url{https://wwwmpa.mpa-garching.mpg.de/gadget4/}}, which uses the same $\mathrm{d}t_a$ definition as $\texttt{AREPO}$ \cite{GADGET4}. This default value is not guaranteed to provided converged results, but our preliminary investigations find that it is sufficient for CDM simulations of our halos. The $\eta$ parameter is one of the three numerical parameters we vary in this work.

The SIDM timestep in \texttt{Arepo} is designed to limit the SIDM scattering probability per timestep to a chosen threshold. This is meant to avoid multiple scatterings within a timestep, which would raise issues of time-ordering the scattering events, as well as provide computational challenges in processor communication in parallel computing environments. \texttt{Arepo} implements SIDM timesteps as:

\begin{equation}
    \delta\mathrm{t}_{\mathrm{SIDM}} = \frac{\texttt{DtimeFac}}{\bar{\rho}\times(\sigma/m)\times\sigma_v}
    \label{eqn:dtsidm}
\end{equation}
where $\sigma_v$ is the local velocity dispersion, $\bar{\rho}$ is the local matter density, $\sigma/m$ is the self-interaction cross-section per unit mass, and \texttt{DtimeFac} is a normalization parameter that we set to 0.0025. \texttt{DtimeFac} is the mean probability for a particle to scatter within the time interval $\delta t_{\mathrm{SIDM}}$, so adhering to this timestep sets a rough limit on the scattering probability any particle will experience during a timestep. Our default value limits this probability to 0.25\%, but in late stages of core-collapse we relax this restriction to allow the probability to reach up to roughly 2.5\%. The local matter density $\bar{\rho}$ is measured for each particle by locating the 32 nearest neighbours, and calculating the density of the sphere centered on the test particle that encloses all 32. The original SIDM module of \texttt{Arepo} only allows $\delta\mathrm{t}_{\mathrm{SIDM}}$ to drop so far, holding it constant after a minimum value is reached \cite{vogelsberger2012}. We have modified our version to continue with increasingly small timesteps, allowing our simulation to reach more advanced stages of core-collapse (see \cite{carton2022} for more details).

In this work we only investigate the influence of the $\eta$ and $r_{\mathrm{soft}}$ parameters in core-collapse. A more detailed timestepping analysis, which may consider alternative timestep definitions and the adjustment of \texttt{DtimeFac} (which represents the scattering probability limit), could play a role in numerical convergence and is a topic for future investigation. An additional parameter we do not vary is the numerical parameter that controls the gravitational tree opening angle (\texttt{ErrTolForceAcc}), for which we use a value of $10^{-3}$.


\subsection{Convergence Testing Grid}
\label{sec:CTgrid}
To rigorously test core-collapse in \texttt{Arepo}, we design a grid of physical and numerical parameter values to simulate. We use the NFW profile, a cuspy halo defined as $\rho_\mathrm{NFW}(r)=\rho_s/(\frac{r}{r_s}(1+\frac{r}{r_s})^2)$, for our initial density profile, as this has shown to fit CDM simulations well \cite{NFW96}. An NFW initial condition with SIDM has three physical parameters to adjust: virial mass ($M_{200m}$), concentration ($c_{200m}$), and the self-interaction cross-section per unit mass ($\sigma/m)$. The subscript $200m$ here represents the mass and concentration definition used, where $M_{200m}$ is defined as the mass enclosed within the spherical region where the average density is 200 times cosmological mean mass density, and $c_{200m}$ is the concentration within that same region. To limit the degrees of freedom in our simulation grid, we fix the halo virial mass at $M_{200m} = 10^{10.5}\: M_\odot$, which is typical of low-mass field dwarf irregulars \cite{mcgaugh2017,Li2019}. An extension of this study to different $M_{200m}$ values may be tested in a future work, in particular lower masses which are characteristic of subhalos.

Fixing the mass leaves concentration and cross-section as our adjustable parameters. To ensure our results are robust to a wide variety of halo structure, we test the halo concentrations of 10 and 50, which cover two ends of a large concentration range. Our halo mass of $M_{200m}=10^{10.5}\: M_\odot$ has a median predicted halo concentration $c_{200m}=19$ at $z=0$ \cite{Diemer2019}, which is contained within our tested interval.

Now that the mass and concentration are accounted for, our final variation is in the scattering cross-section. Instead of using two fixed values of $\sigma/m$ in our halo grid, we instead adjust the cross-section as needed in order to produce two fixed values of collapse time. We make this choice because our preliminary testing indicated there may be numerical errors that depend on simulation length, rather than the value of $\sigma/m$. These two concentrations and two collapse times produce four combinations of physical parameters, which can also be represented by their four unique values of $\hat{\sigma}$, falling between $0.00827$ and $7.03$. These values sample across the smfp and lmfp regime transition. The three simulations falling in the lmfp regime will make our findings valuable for $\beta$ calibration, which must be done to establish a consistent heat transfer model that can be used for all values of $\hat{\sigma}$. The one simulation in the smfp regime is useful as a test of the N-body simulator, which should produce valid results regardless of the scattering regime.

To tune the cross-section to any particular collapse time, we use the following prediction:

\begin{widetext}
\begin{equation}
    t_c =
    \begin{cases}
        \frac{150}{\beta}\frac{1}{\hat{\sigma}}\frac{1}{\sqrt{4\pi G \rho_s}} & \text{if } \hat{\sigma} < 0.1\\
        3.58\times10^8 \left(\frac{\sigma/m}{\mathrm{cm}^2/g}\right)^{-0.74}
        \left(\frac{M_{200m}}{M_\odot}\right)^{-0.24}c_{200m}^{-2.67}
        \mathrm{Gyr}& \text{if } \hat{\sigma} > 0.1.\\
    \end{cases}
\end{equation}
\end{widetext}

The first collapse prediction is equation 4 in \cite{essig_SIDM}, which has been calibrated to halos with $\hat{\sigma} < 0.1$. In our parameter selection we use a fiducial $\beta$ value of 1.0. The precise $\beta$ value is not important since the collapse time predictions do not need to be precise, as long as the true collapse times lie in the two desired regimes. The second case is equation 9 in \cite{carton2022}, which has been calibrated to $\hat{\sigma} > 0.1$. We find that the predicted collapse times reasonably match the actual collapse times observed in the most converged simulations.

\begin{table}
	\centering
	\caption{The four simulation batches used in the \texttt{Arepo} convergence testing. This table lists the name, initial concentration, estimated collapse time, scattering cross-section, and $\hat{\sigma}$ for each batch.}
	\label{tab:scenario_table}
	\begin{tabular}{lcccc} 
		\hline
		Simulation & $c_{200m}$ & $t_c$ (Gyr) & $\sigma/m$ ($\mathrm{cm}^2/g$) & $\hat{\sigma}$\\
		\hline
        \hline
        C10T9 & 10 & 9 & 1892 & 7.03\\
        \hline
        C10T225 & 10 & 225 & 17.65 & 0.0656\\
		\hline
        C50T9 & 50 & 9 & 5.688 & 0.267\\
        \hline
        C50T225 & 50 & 225 & 0.1764 & 0.00827\\
		\hline
	\end{tabular}
\end{table}

Using this collapse time prediction, we calculate four cross-sections by pairing the two halo concentrations with the two collapse times of 9 Gyr and 225 Gyr. These choices of collapse time, like those for concentration, represent the extreme ends of a large range of values. The 9 Gyr collapse time will result in a halo that undergoes significant gravothermal collapse within the age of the universe, while the 225 Gyr case results in much more gradual evolution. The 225 Gyr case is also chosen to have a similar collapse time as the NFW halo tested in \cite{koda2011}, which has been the basis of recent $\beta$ calibration \cite{essig_SIDM}. The four physical scenarios generated from the concentration and collapse time choices are shown in Table \ref{tab:scenario_table}, and throughout this work we identify the combinations of physical parameters using their concentration value and estimated collapse time in Gyr as shown in the leftmost column of Table \ref{tab:scenario_table}.

\begin{figure}
\begin{subfigure}[t]{\columnwidth}
    \centering
	\includegraphics[width=\textwidth]{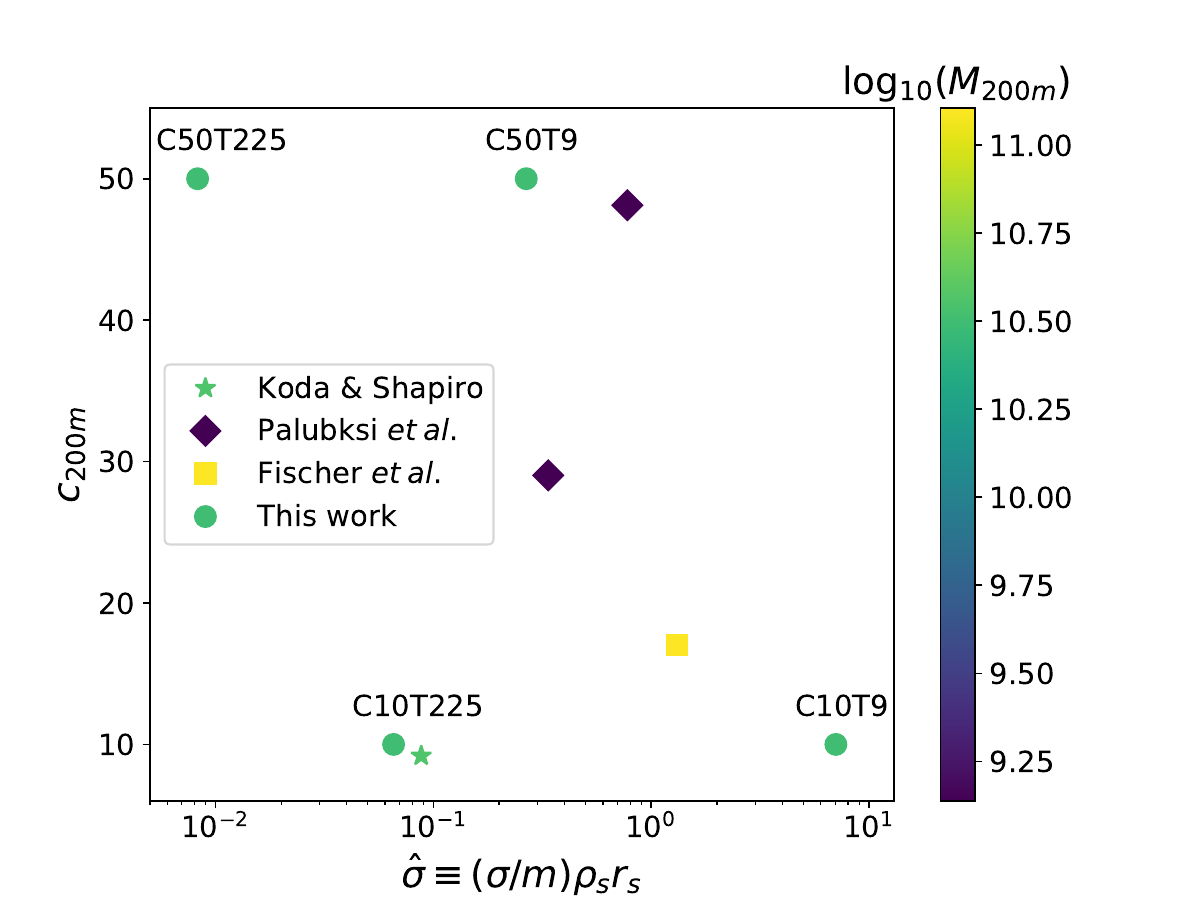}
\end{subfigure}
    \caption{A comparison of physical halo parameters used in this work to other studies\cite{koda2011,palubski2024numerical,fischer2024numerical}. The four physical cases in this work are identified by the names listed in Table \ref{tab:scenario_table}. $M_{200m}$, $c_{200m}$, and $\hat{\sigma}$ for the three other works were calculated using their reported initial $r_s$ and $\rho_s$ values.}   \label{fig:CT_regimes}
\end{figure}

Figure \ref{fig:CT_regimes} shows a comparison of our four physical parameter choices to other studies in the literature. We compare to the NFW halo simulation in Koda \& Shapiro's gravothermal fluid study \cite{koda2011}, as well as the recent publications by Palubski \emph{et al.} \cite{palubski2024numerical} and Fischer \emph{et al.} \cite{fischer2024numerical} that investigate numerical errors in core-collapse simulations. Our study spans a wide range in $\hat{\sigma}$, which allows us to show how numerical errors behave in different scattering regimes. The studies by Palubski \emph{et al.} and Fischer \emph{et al.} probe relatively low and high mass halos respectively, as well as the halo concentration range within our parameter grid. Our C10T225 halo has physical parameters comparable to the Koda \& Shapiro's simulation, which has been used to calibrate the gravothermal fluid $\beta$ parameter.

For each of these four physical scenarios we run 36 simulations, varying over a grid of values in the non-physical parameters particle number ($N$), timestep accuracy ($\eta$), and softening length ($r_{\mathrm{soft}}$). For $N$ we use the values $10^4$, $10^{4.25}$, $10^{4.5}$, $10^{4.75}$, $10^5$, and $10^6$. The closely packed values from $10^4$ to $10^5$ give a detailed picture of the resolution dependence, while the jump to $10^6$ tests whether or not the $10^5$ results are reasonably converged. For $\eta$ we use the default value of 0.025, and extend downwards by factors of 5 to 0.005 and 0.001. The $r_{\mathrm{soft}}$ values are based on a standard softening length, depending on particle count $N$ and halo concentration $c$:
\begin{equation}
    \epsilon(N,c) = r_s\left[\ln{(1+c) - \frac{c}{1+c}}\right]\sqrt{\frac{0.32(N/1000)^{-0.8}}{1.12c^{1.26}}}.
    \label{eqn:soft}
\end{equation}
This standard softening value is designed to be large enough to avoid spurious two-body gravitational scattering and small enough that the gravitational force is not over-smoothed in CDM simulations \cite{vdB2018,carton2022}\footnote{This particular softening criterion was developed for subhalos, not the isolated halos simulated in this work. Since we will be varying the softening length over an order of magnitude in this investigation, we find this softening assignment to be an acceptable starting point.}. We determine this standard value for each resolution and concentration combination, and then vary the softening using fixed multipliers on $\epsilon$. The values we test are $10^{-0.4}\epsilon(N,c)$, $\epsilon(N,c)$, $10^{0.4}\epsilon(N,c)$, and $10^{0.8}\epsilon(N,c)$.

For each value of resolution and concentration, we generate initial conditions using the \texttt{SpherIC} code \citep{spherIC}. The initial particle positions and velocities are drawn from an NFW distribution with an isotropic velocity dispersion. To keep the halo mass finite, our initial density profile is truncated at the virial radius using an exponential cutoff as presented in \cite{halogen4muse}. Core-collapse occurs in a relatively small central region compared to the halo's virial radius, so we expect that the details of our truncation will not affect our results \cite{shengqi2023}. The manner of truncation may be more important for simulations of non-isolated halos, which may be tidally stripped to smaller radii \cite{carton2022,carton2023}.

\subsection{Analysis}

To calculate the central mass density for a given snapshot, we adopt the method used in \cite{koda2011}. This calculation assumes the profile within the core radius $r_c$ is an isothermal sphere, and finds $r_c$ iteratively using the constraint:

\begin{equation}
    r_c=\sqrt{v_c^2/4\pi G\rho_c}
    \label{eqn:CD_cond}
\end{equation}
with $v_c$ as the velocity dispersion inside $r_c$, and the density $\rho_c$ defined as a function of the mass $M(r_c)$ enclosed within $r_c$:
\begin{equation}
    \rho_c\equiv 1.10\times M(r_c)\left/\frac{4}{3}\pi r_c^3\right. .
\end{equation}
 The factor of 1.10 is the ratio between the central density and the average density within $r_c$, and can be found in \cite{koda2011}. To calculate $r_c$ we begin with an initial guess and use that value to compute $v_c$ and $\rho_c$. We then use those values in Equation \ref{eqn:CD_cond} to find a new $r_c$, and iterate until the radius converges.  This method uses as many particles as possible to calculate the central density without extending beyond the core radius. Methods using fixed numbers of particles risk underestimating the density by using a core radius that is too large, and require the number of particles being considered to be fine-tuned. A downside of our method is that, for low-resolution halos, a solution to Equation \ref{eqn:CD_cond} is not guaranteed. When this occurs, the simulation resolution is too low to consistently resolve the core regardless of definition. We found in our analysis that nearly all snapshots had solutions to Equation \ref{eqn:CD_cond}. All exceptions were for $10^4$ particle simulations, none of which missed enough points that the central density evolution was unclear. In figures, missing points are filled by interpolating between the neighboring points, using an exponential function that is linear in the semi-log plotting space.

To measure the collapse time of a simulation, we first smooth the central density evolution. We use a simple smoothing method, where each central density measurement is averaged with the two adjacent points in time. We then find the time at which this smoothed central density evolution first exceeds $20\rho_s$. This threshold is chosen because it is far enough along the core-collapse timeline that simulations with different collapse rates are distinguishable from one another, and after this point they all evolve similarly as the collapse rate sharply increases. Using a higher value would require additional computation time as scattering events become more frequent. We found that adjusting the central density threshold by a factor of order unity did not substantially alter the relations found in this study. Our method is more direct than other collapse time measurements in the literature, which measure it based on the core's relaxation time \cite{koda2011} or the Knudsen number \cite{essig_SIDM}. We use the collapse time to compare numerical parameter choices within this work, not to compare with other works, so this difference in definition does not limit our analysis.

\section{Realization Noise}

\label{sec:noiseTests}

\begin{figure*}
    \begin{subfigure}[t]{\textwidth}
	\includegraphics[width=\columnwidth]{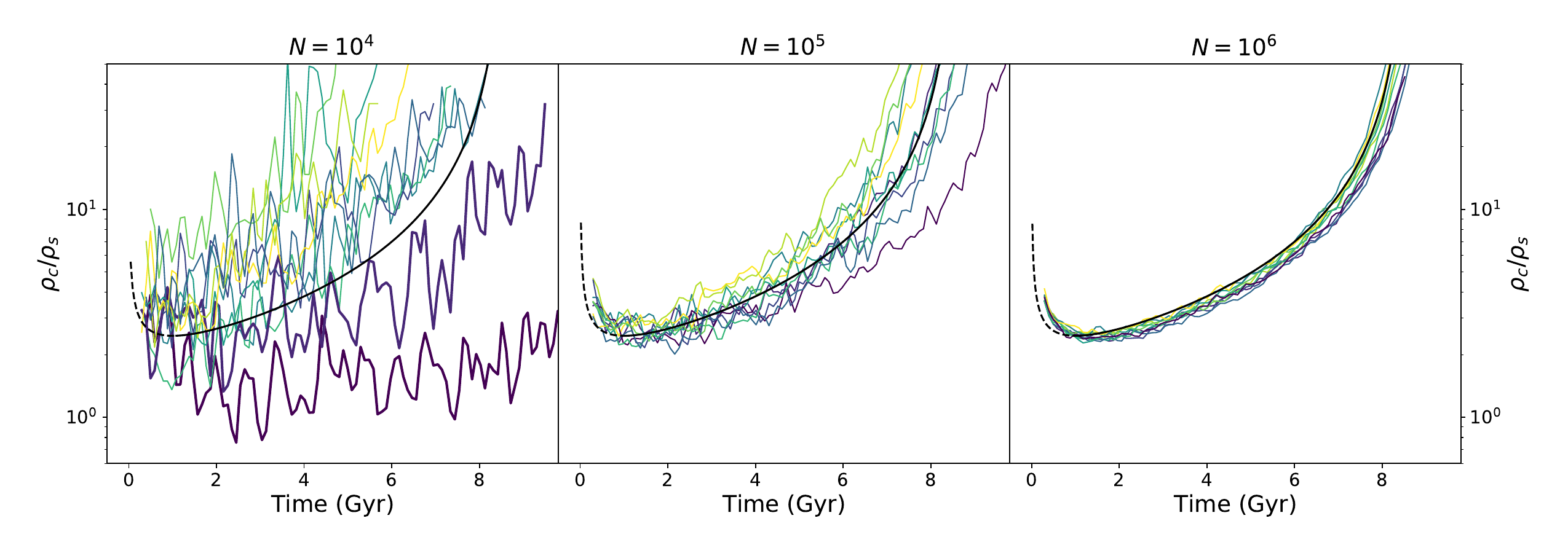}
    \caption{}   
    \end{subfigure}
    \begin{subfigure}[t]{\textwidth}
        \includegraphics[width=\columnwidth]{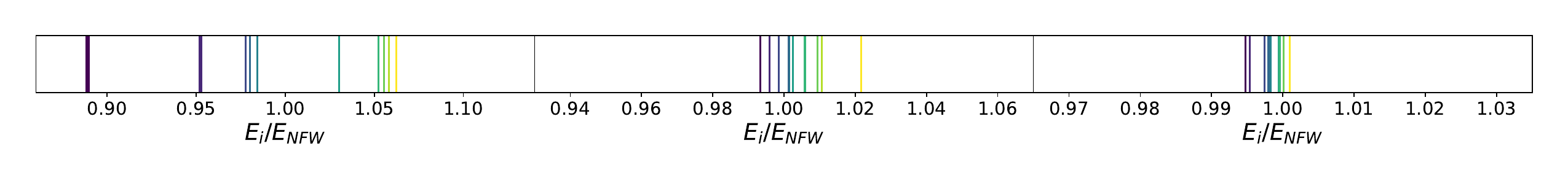}
        \caption{}
    \end{subfigure}
    \caption{Realization noise tests for C50T9 ($\hat{\sigma}=0.2664$) at three different resolutions. \textbf{a)} The central density evolution. The bold black line is the gravothermal model with $\beta=0.9385$, and each other line represents an individual realization. \textbf{b)} The initial total energy of each realization relative to the expected NFW total energy.  Each realization has the same line coloring in the top and bottom panels. Two outliers in the $N=10^4$ case are shown in bold in both panels for emphasis.} \label{fig:noise_C50T9}
\end{figure*}

\begin{figure*}
    \begin{subfigure}[t]{\textwidth}
	\includegraphics[width=\columnwidth]{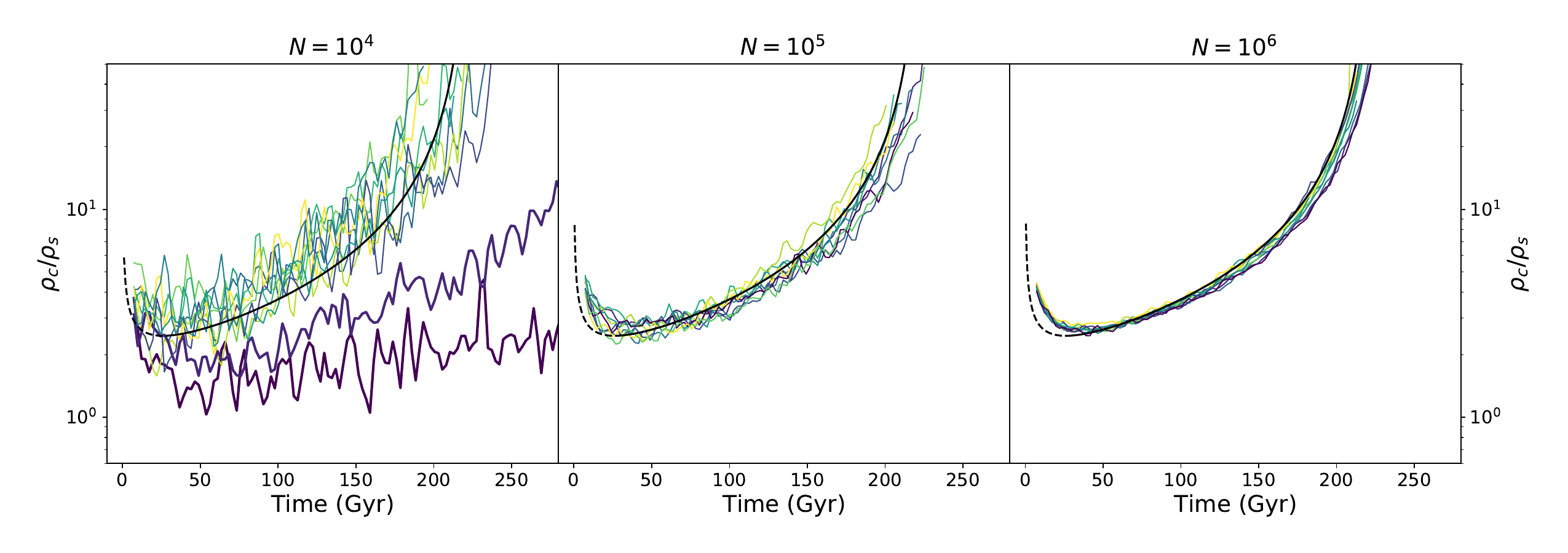}
    \caption{}   
    \end{subfigure}
    \begin{subfigure}[t]{\textwidth}
        \includegraphics[width=\columnwidth]{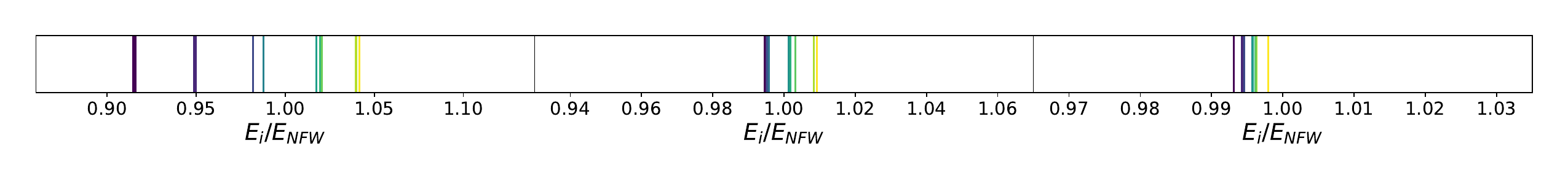}
        \caption{}
    \end{subfigure}
    \caption{Realization noise tests for C10T225 ($\hat{\sigma}=0.0657$) at three different resolutions. Panel organization is the same as in Figure \ref{fig:noise_C50T9}. The bold black line is the gravothermal model with $\beta=1.162$.}\label{fig:noise_C10T225}
\end{figure*}

\begin{figure*}
	\includegraphics[width=2\columnwidth]{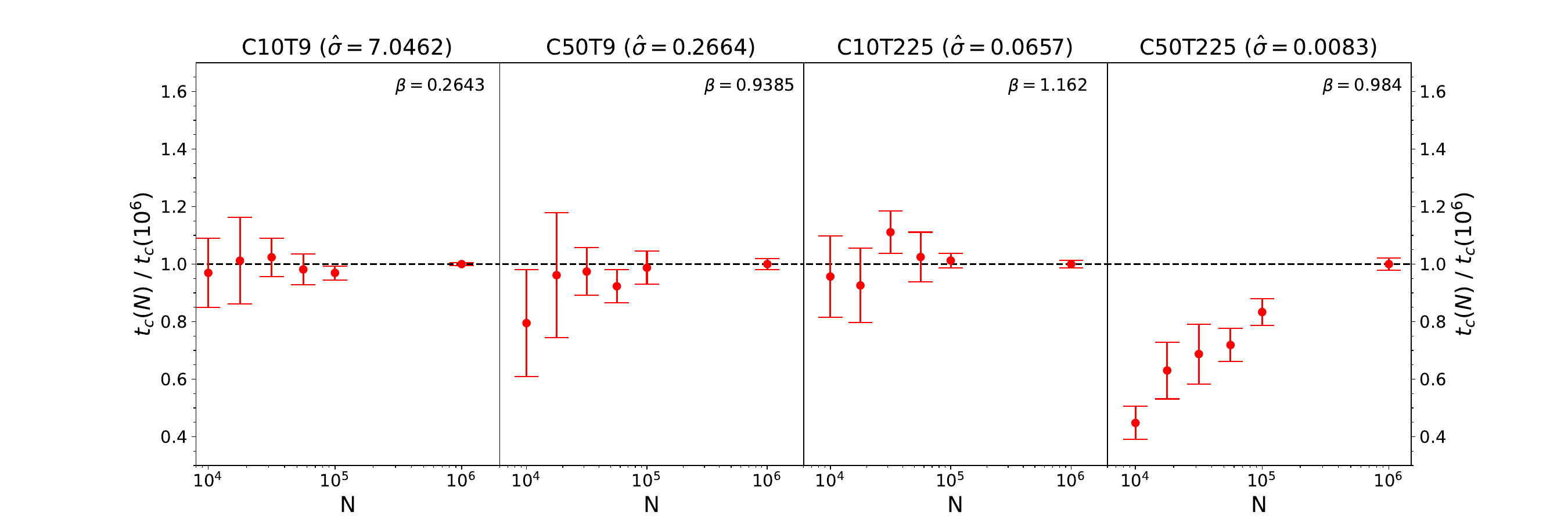}
    \caption{Collapse time statistics as a function of resolution for all realization noise tests. The points are the median collapse time $t_c(N)$ from simulations, scaled by the best-fit collapse time $t_c(10^6)$ from the gravothermal model calibrated to $N=10^6$. The error bars are the median absolute deviation of $t_c(N)$ at each value of $N$. Each panel shows results from one halo in Table \ref{tab:scenario_table}, and displays the best fit $\beta$ for that case in the upper right corner.}   \label{fig:noise_tcplot}
\end{figure*}

Before we can interpret the convergence testing results, we need to understand the effect of realization noise on our simulations. Realization noise is introduced into our simulations in two ways: initial conditions and scattering. The initial condition noise is introduced when we sample an analytical density profile to produce a halo of discrete particles. Sampling using different random seeds, even if the NFW parameters do not change, can produce slightly different distributions of simulation particles. If these fluctuations are large enough they could affect the halo evolution, including the collapse time.

Scattering noise is associated with a different random seed, which is used to generate random numbers used in the SIDM scattering algorithm. These numbers are compared to the scattering probability to determine which particle pairs have scattering events. Different scattering seeds may produce different scattering rates at important stages of core-collapse, which could affect the collapse time.

In this study we do not directly vary the scattering seed, only the initial condition seed. Changing the initial condition seed produces an entirely new set of particles that are sorted onto the processors in a different order. This means that for two different initial condition realizations, the same scattering seed will produce two independent scattering histories. Our results in this section therefore include the effects of both scattering and initial condition noise, but do not study the two separately.

To study realization noise we run multiple simulations with the same softening and $\eta$, but with differently seeded random initial conditions, as a function of the total number of particles in the simulation. We use equation \ref{eqn:soft} for the softening and set $\eta=0.005$. These choices are based on preliminary testing that showed reasonable convergence when these values are selected, and are further justified in section \ref{sec:CT}.

In Figure \ref{fig:noise_C50T9}, we show the results of our  tests for the C50T9 halo. Selecting the C50T9 case shows the behavior of a halo in the lmfp region that is collapsing on a cosmological timescale. We display the evolution of the halo central density divided by the initial NFW scale density ($\rho_c/\rho_s$) for three different numbers of SIDM particles in the initial conditions: $N=10^4$, $N=10^5$, and $N=10^6$. The evolution of $N=10^4$ particle simulations has significant scatter, with typical collapse times varying between 4 and 8 Gyr, and extreme outliers taking even longer to collapse. The $N=10^5$ and $N=10^6$ realizations have less scatter, with the $N=10^5$ collapse times varying between 6 and 8 Gyr and the $N=10^6$ collapse times closely distributed at around 8 Gyr. In Figure \ref{fig:noise_C10T225} we show the same analysis for C10T225, as a slowly collapsing halo with different initial conditions may have a different pattern of realization noise. This figure shows similar results, with the lowest resolution suffering from overwhelming realization noise and extremely slowly collapsing outliers. Similar analyses of C10T9 and C50T225 find that these features are generic to all initial conditions and cross-sections that we have tested.

In both Figures \ref{fig:noise_C50T9} and \ref{fig:noise_C10T225}, the initial energies are typically below the expected value at the highest resolution. The size of the offset is smaller than 1\%, and is likely not a concern for this study. This small discrepancy may be due to changes to the potential energy introduced by gravitational softening \cite{barnes2012}.

From these examples we see that $10^4$ particles is insufficient to consistently resolve the core-collapse process in this case. In addition, the $N=10^4$ cases suffer from a small number of slowly collapsing outliers and a cluster of realizations with slightly accelerated evolution. While there is collapse time scatter in the $N=10^5$ realizations, they are more evenly distributed around the expected evolution.

The slowly collapsing $N=10^4$ outliers in C50T9 and C10T225 have one characteristic in common: these simulations all have initial total energies (defined as the summed kinetic and potential energies of all simulation particles) that are lower than all of the other realizations, as shown in the bottom left panels of Figures \ref{fig:noise_C50T9} and \ref{fig:noise_C10T225}. Since the total energy of the system is negative, $E_i/E_{NFW}<1$ implies $E_i>E_{NFW}$. This indicates a halo with less tightly bound particles, either from a less concentrated halo (as particles are spread out, the negative gravitational potential energy gets larger and approaches zero), or from a halo with higher velocity particles. In this work the outliers are caused by low magnitude values of the gravitational potential energy, not large kinetic energies. It is possible that the particles in these realizations are slightly less concentrated than ideal NFW case, and consequently take longer to experience core-collapse.

To explore the bias in collapse time as well as the scatter, we show both as a function of particle number for all our halos in Figure \ref{fig:noise_tcplot}. This figure includes the other two halos in Table \ref{tab:scenario_table}, and additional resolutions between $10^4$ and $10^5$. The C10T9 halo, with $\hat{\sigma}=7.0462$, is in the smfp regime and therefore is not well represented by the gravothermal fluid model discussed in section \ref{sec:method}. Despite this, we fit the collapse time and present the associated value of $\beta$. The absolute value of this fit is less physically meaningful than those for the other three halos, but still provides a useful reference point to analyze collapse time scatter. Each data point and associated error bar in Figure \ref{fig:noise_tcplot} represents 10 realizations of that halo at the specified resolution. To minimize the effect of the extreme outliers at low resolution, the points in this figure are the median collapse time and error bars are the median absolute deviation. For most cases, the median collapse time is consistent across resolutions and is more tightly distributed as the resolution increases. There are, however, important exceptions to this trend.

Halos C50T9 and C50T225 have consistently short collapse times at the lowest resolution. This suggests that collapse times in low resolution halos in this case are not only noisy, but are not centered on the converged result. The collapse time of halo C50T9 converges by $N=10^{4.25}$, but the collapse times of halo C50T225 increase through all of the resolutions we have tested. This suggests that the C50T225 halo has not converged by $N=10^5$, and may not be converged even at $N=10^6$. Higher resolution tests of C50T225 are numerically expensive, and the halo may be too difficult to resolve to be of use in simulation work until this problem is remedied.

\begin{figure}
\begin{subfigure}[t]{\columnwidth}
    \centering
	\includegraphics[width=\textwidth]{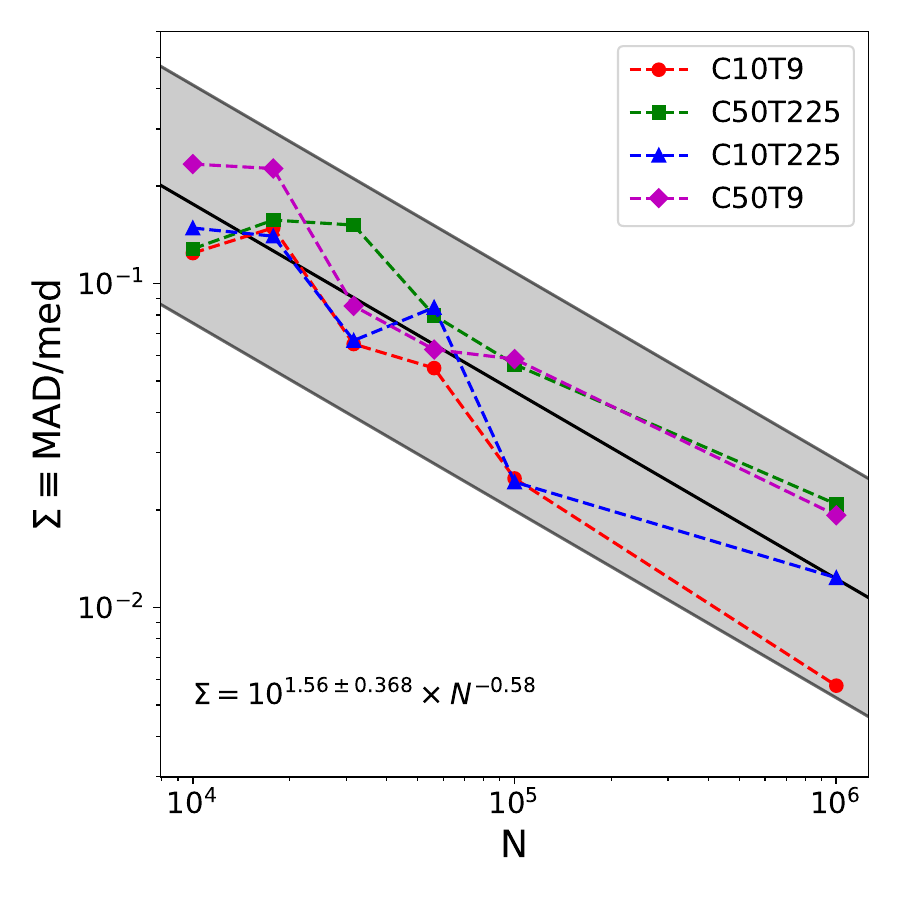}
\end{subfigure}
    \caption{The fractional scatter of the collapse times shown in Figure \ref{fig:noise_tcplot} as a function of $N$. Different scenarios from Table \ref{tab:scenario_table} are distinguished by color and marker style. The power law fit given in the lower left corner is shown by the central black line, and the uncertainty in the normalization constant is represented by the shaded region.}   \label{fig:tc_scatter}
\end{figure}

The last step before we can interpret our convergence testing results is to determine to what precision we can measure collapse time for a given simulation. In Figure \ref{fig:tc_scatter} we show the error bars, $\Sigma$, used in Figure \ref{fig:noise_tcplot}, defined as the median absolute deviation divided by the median collapse time, as a function of particle count. From these data we see that the collapse times in our highest resolution simulations can be considered to have a precision of 3\%, while our lowest resolutions can have uncertainties as high as 20\%. We also find that the collapse time scatter is well described by the power law $\Sigma\propto N^{-0.58}$ with the constant of proportionality for our work depending somewhat on the specific scenario, ranging from $10^{1.192}$ to $10^{1.928}$. The exponent on $N$ and the normalization exponent were both found by averaging the values found when fitting each scenario individually, and the range in the normalization is the standard deviation from those individual values.

From these data we have found that simulations with $10^4$ particles suffer from significant realization noise, to the point where any individual $N=10^4$ halo cannot be a reliable representation of a larger population of similar halos, and even an ensemble may have a biased average collapse time. Additionally, we have found that $N=10^4$ halos are sensitive to the initial total halo energy, with low energy magnitudes correlating with collapse times much longer than the rest of the population. We have also found currently unexplained systematic effects at low resolutions, with the extreme case of C50T225 not converging in collapse time by a resolution of $N=10^5$. All of these findings suggest that a halo must have at least $10^5$ particles to produce a reliable collapse time. Lower resolutions have collapse time uncertainties rising above 10\%, extremely slowly collapsing outliers, and in some cases a median collapse time that disagrees with the converged result. Our data also suggest that C50T225, our high concentration and low cross-section case that is firmly in the lmfp regime with $\hat{\sigma}=0.0083$, suffers from further numerical error that persists until higher resolutions than $N=10^5$ with our current softening and $\eta$ choices.

\section{Convergence Testing}

\label{sec:CT}

\begin{figure*}
	\includegraphics[width=2\columnwidth]{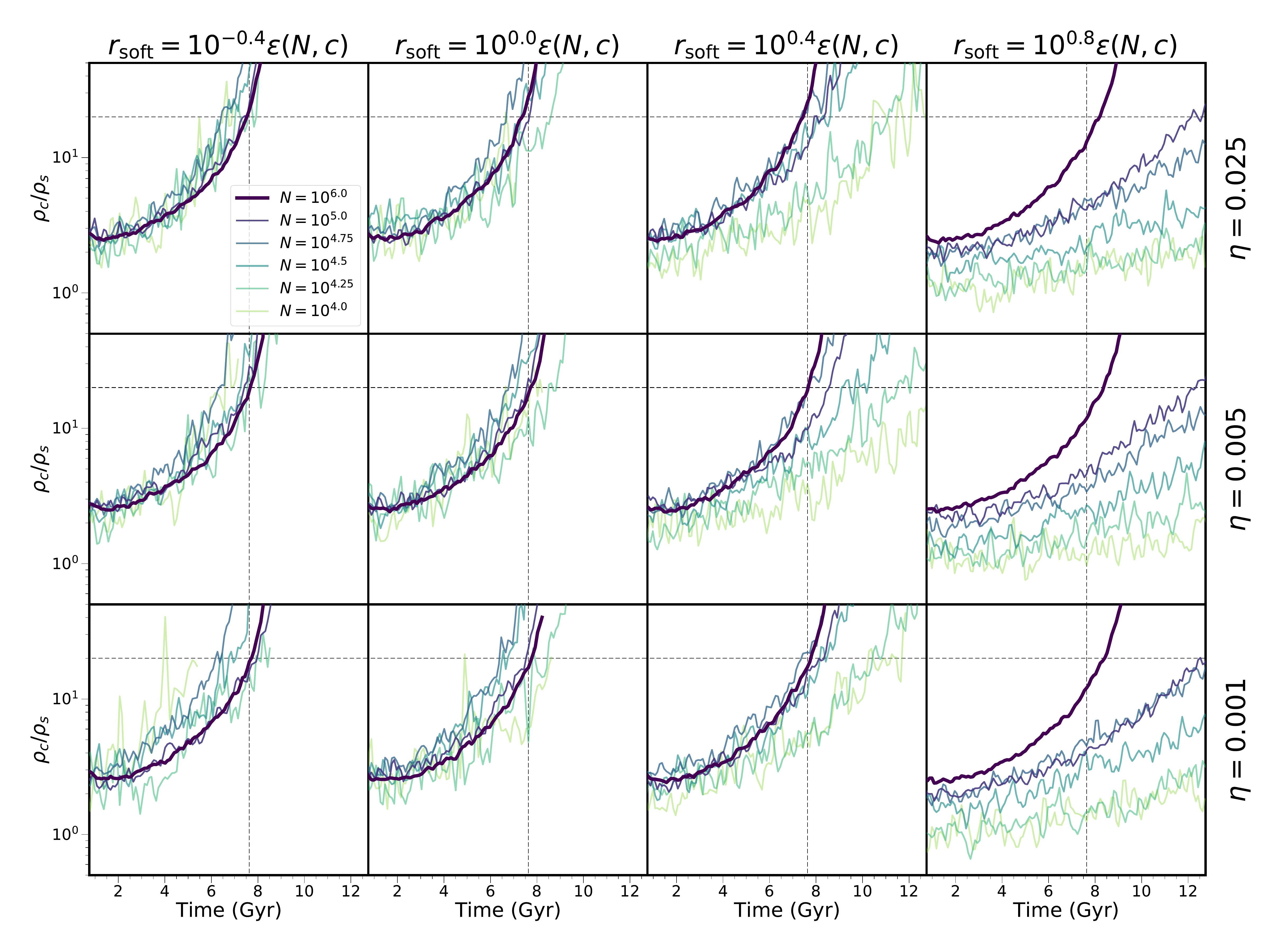}
    \caption{Convergence testing results for halo C50T9. Each panel shows the central density evolution at various resolutions. Panels in the same column share the same softening parameter, and panels in the same row share the same $\eta$, as labeled on the top and right sides of the figure respectively. The $N=10^6$ simulation line is bold in each panel. The horizontal dashed lines mark $\rho_c/\rho_s=20$, our core-collapse threshold, and the vertical dashed lines mark the core-collapse time of our $r_{\mathrm{soft}}=\epsilon(N,c)$, $\eta=0.001$, $N=10^6$ case.}   \label{fig:gridC50T9}
\end{figure*}

\begin{figure*}
    \centering
	\includegraphics[width=2\columnwidth]{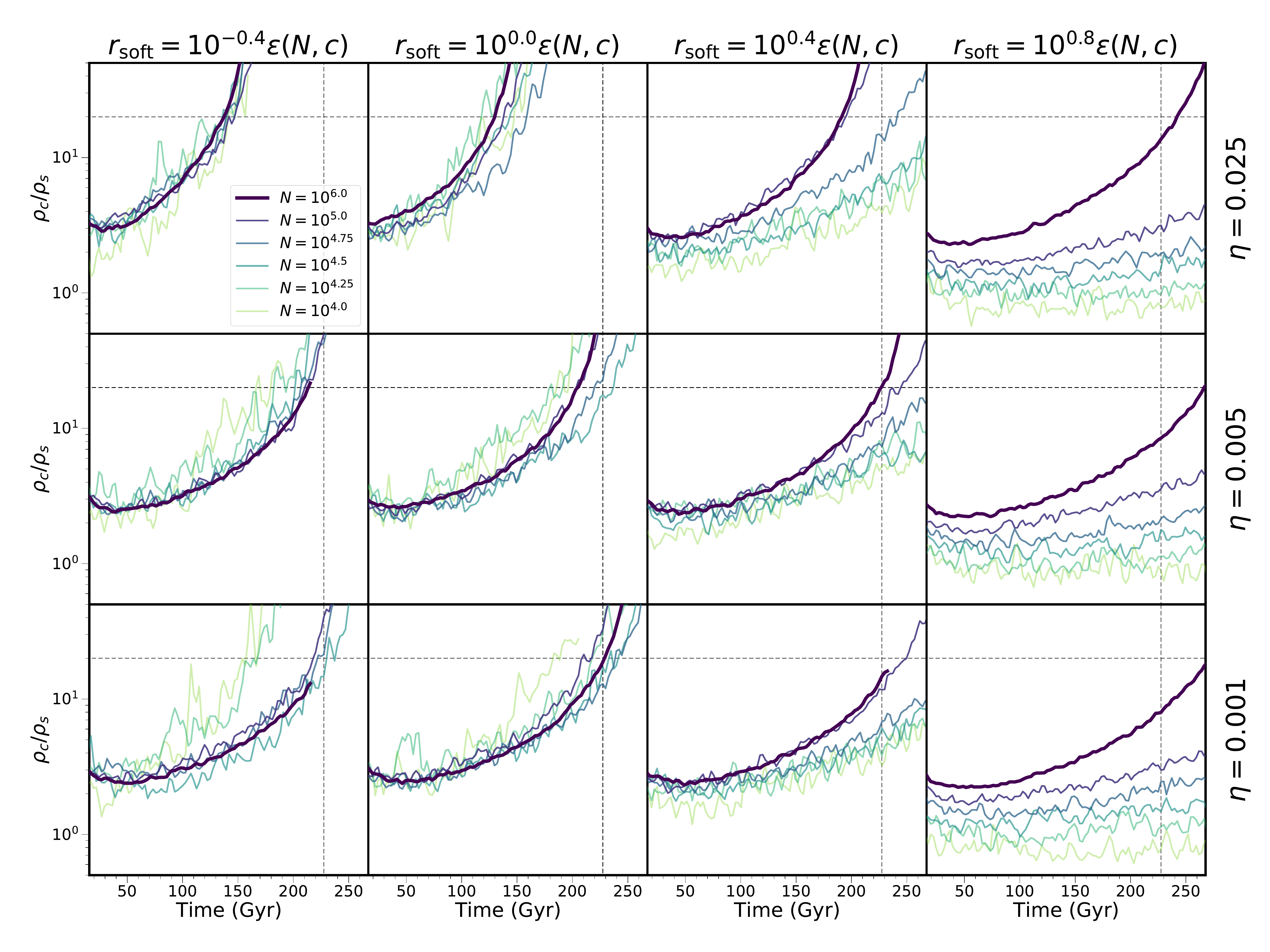}
    \caption{Convergence testing results for halo C10T225. All lines and markers are the same as in Figure \ref{fig:gridC50T9}.}   \label{fig:gridC10T225}
\end{figure*}

\begin{figure*}
	\includegraphics[width=2\columnwidth]{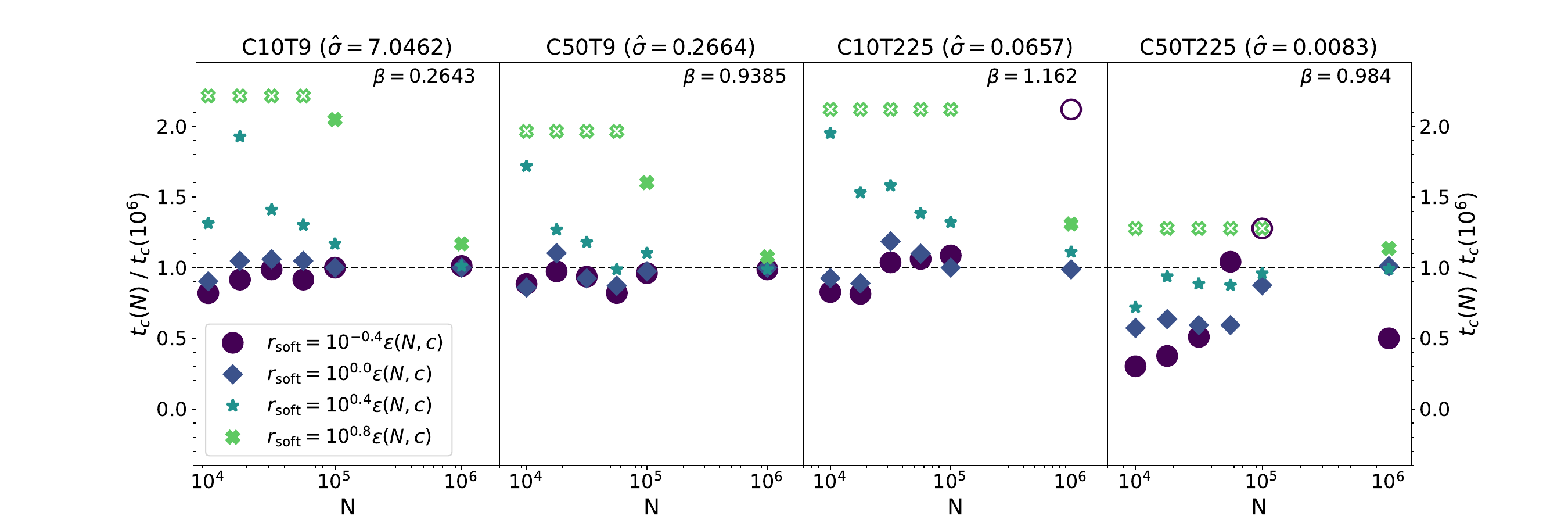}
    \caption{Collapse time as a function of resolution and softening length, with $\eta=0.005$. Each panel shows a different scenario from Table \ref{tab:scenario_table}, and different softening lengths are distinguished by color and marker style. In the same manner as Figure \ref{fig:noise_tcplot}, the horizontal dashed line represents the expected collapse time using $\beta$ values fit in Section \ref{sec:noiseTests}, which are displayed in the top right corner of each panel. Unfilled markers represent the lower bound on collapse time for simulations that never reached the core-collapse threshold.}   \label{fig:resolutionTest}
\end{figure*}

\begin{figure*}
	\includegraphics[width=2\columnwidth]{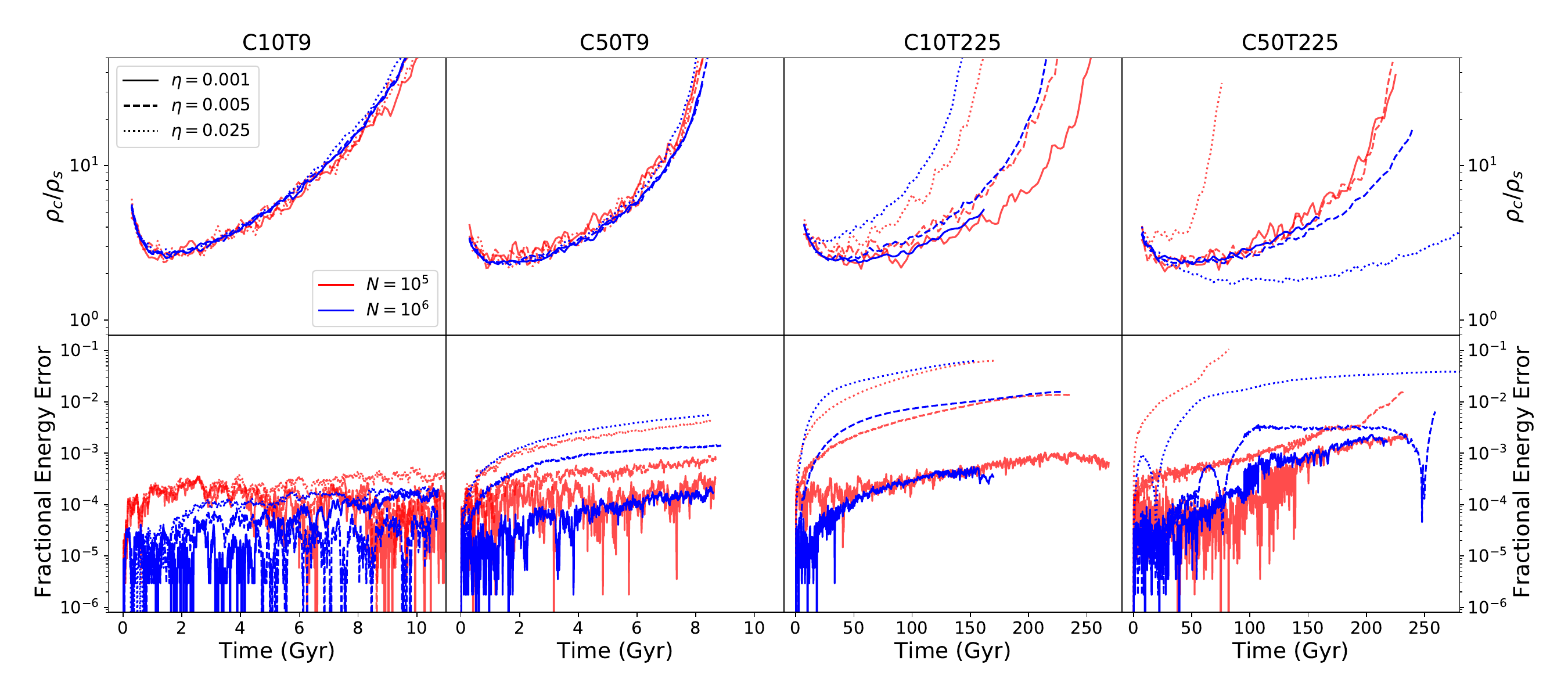}
    \caption{Central density and absolute fractional energy error evolution as a a function of $\eta$, shown for $N=10^5$ (red) and $N=10^6$ (blue). Each column shows a different scenario from Table \ref{tab:scenario_table}, with the top panels displaying the central density evolution and the bottom panels the energy error evolution. Each panel shows $\eta=0.001$ (solid), $\eta=0.005$ (dashed) and $\eta=0.025$ (dotted). All simulations shown use $r_{\mathrm{soft}}=\epsilon(N,c)$ as defined in Equation \ref{eqn:soft}.}   \label{fig:eta_evolution}
\end{figure*}

Having quantified the realization noise, we can analyze the main convergence test results following the grid described in section \ref{sec:CTgrid}. First we show the central density evolution across the numerical parameter grid for two sample halos and identify important variation in the collapse time as a function of $r_{\rm soft}$ and $\eta$ individually. Subsequently, we consider summary data, to further quantify these trends across all halos.

Figures \ref{fig:gridC50T9} and \ref{fig:gridC10T225} show convergence testing results for halos C50T9 and C10T225 respectively, the same halos shown in Figures \ref{fig:noise_C50T9} and \ref{fig:noise_C10T225}. The central density plots are arranged in a grid according to their softening length and $\eta$ values, and different resolutions are distinguished by color. Following our guidance from section \ref{sec:noiseTests}, when generating $N=10^4$ initial conditions we have required that the initial total energy is within 5\% of the expected analytic NFW energy. This prevents our analysis from including an extremely slowly collapsing outlier.

We first consider the effect of softening on convergence. We use Equation \ref{eqn:soft} as the default value, $\epsilon(N,c)$, of the softening length, and test softenings scaled up and down from this default. In these tests, oversoftening manifests as a delayed core-collapse which is more significant at lower resolutions. This can be seen in both Figures \ref{fig:gridC50T9} and \ref{fig:gridC10T225}, where large softening values (the two rightmost columns) show delayed central density evolution even for large particle counts. This effect occurs in both simulations, but is more severe for C10T225.  On the other hand, under-softening leads to the opposite effect, but only for low particle resolution. Due to the higher magnitude of error and prevalance at even the highest resolutions, oversoftening is a much more severe problem for convergence than undersoftening.

To quantify the effect of softening on convergence, Figure \ref{fig:resolutionTest} shows a more detailed comparison of softening, resolution, and collapse time, and includes the C10T9 and C50T225 data. Each panel shows the collapse time as a function of particle count for one of our halos from Table \ref{tab:scenario_table}, and the different symbols and colors show softening length choices. All simulations represented in this figure are for $\eta=0.005$. Some simulations did not core-collapse in the time given, and are represented as unfilled markers at the end of their run. The maximum simulation time varies between panels based on how numerically expensive the simulations are, as well as the expected collapse time value. For all halos except C50T225, the collapse time at high resolution converges reasonably for $r_{\mathrm{soft}}$ values of $10^{-0.4}\epsilon(N,c)$, $10^{0}\epsilon(N,c)$, and $10^{0.4}\epsilon(N,c)$. Simulations with $10^{0.8}\epsilon(N,c)$, our most oversoftened cases, produce consistently delayed collapse even at $N=10^6$. The standard softening choice ($10^{0}\epsilon(N,c)$) and the undersoftening value ($10^{-0.4}\epsilon(N,c)$) appear to be largely insensitive to resolution for $N>10^4$, while the moderate oversoftening case ($10^{0.4}\epsilon(N,c)$) displays delayed collapse at resolutions up to and including $N=10^5$. Moderate undersoftening ($10^{-0.4}\epsilon(N,c)$) produces similar collapse times as the standard softening in all cases except C50T225. The C50T225 halo does not display any of the trends discussed above, and the different softening values do not produce consistent results at our highest tested resolution. The standard softening case suffers from accelerated collapse at low resolutions. This accelerated collapse is consistent with the C50T225 panel in Figure \ref{fig:noise_tcplot}, where we observed the same effect. Surprisingly, the moderate oversoftening case has the least variation with resolution in the C50T225 simulations, with resolutions above $N=10^4$ collapsing near the expected rate.

Next, we consider the effect of the choice of the time step size for convergence.  Here, we focus on the default time step parameter $\eta$, where smaller values of $\eta$ produce smaller timesteps for all particles as shown in equation \ref{eqn:dta}. For the long collapse time results in Figure \ref{fig:gridC10T225}, larger values of $\eta$ cause the central regions of the halo to collapse more rapidly than with smaller values. This is most clear in the second column from the left, where the standard softening length is used. Decreasing $\eta$ from the default value of 0.025 to 0.005 increases the core-collapse timescale by roughly 40\%, while further decreasing $\eta$ to 0.001 results in a smaller additional increase, bringing the total increase to 65\%. The problem is less severe for shorter, cosmological timescales. Figure \ref{fig:gridC50T9} shows no sign of this $\eta$ dependence, with $\eta=0.025$ appearing sufficient for a converged core-collapse time.

Some odd behavior appears at small softening (leftmost column) in Figures \ref{fig:gridC50T9} and \ref{fig:gridC10T225}, in particular for the lowest resolution $N=10^4$. These simulations agree with the converged result for large $\eta$, but have accelerated collapse times for smaller $\eta$. In Figure \ref{fig:gridC50T9} this effect is small and vanishes at $N=10^{4.25}$, so it is easily attributed to the realization noise shown in Section \ref{sec:noiseTests}. Figure \ref{fig:gridC10T225} shows a stronger trend however, with the $N=10^{4.25}$ simulation also experiencing accelerated collapse at small $\eta$. Even with this stronger trend, we suspect the realization noise is the dominant effect in this counterintuitive variation. Additional realizations for each of these $\eta$ values ware required to distinguish between a shot-noise effect and any true dependence on $\eta$ at small softening and resolution, and this may be explored in future work.

Figure \ref{fig:eta_evolution} shows how the choice of $\eta$ influences the halo evolution. Each column shows a different halo from table \ref{tab:scenario_table}, with the top row plotting the central density evolution and the bottom row plotting the energy error. Different resolutions are represented by the line color, while $\eta$ values are distinguished by the line style. The central density evolution of the quick collapse cases, C10T9 and C50T9, have no significant dependence on $\eta$ in the range we tested. In contrast, the slow collapse cases show strong dependence on the choice of $\eta$, with smaller values leading to longer core-collapse times. In both of the slow collapse cases the default value of $\eta=0.025$ is insufficient, with collapse times differing from the converged case by as much as a factor of 4, but even the smallest value of $\eta=0.001$ may not be sufficient in all cases. The C10T225 simulation has a longer collapse time when using $\eta=0.001$ compared to $\eta=0.005$ at both resolutions shown, which means an $\eta$ smaller than $0.001$ may be needed for convergence. The C50T225 case fares better than C10T225, with $\eta=0.005$ and $\eta=0.001$ showing similar evolutions. The $N=10^5$ data do not match the $N=10^6$ data however, suggesting higher resolution sensitivity than the short collapse simulations.

The $\eta$ dependence in the central density evolution is tracked well by the energy error, with larger $\eta$ values accumulating larger errors. The C10T9 case shows small amounts of energy error accumulation, with most energy errors being dominated by random noise rather than secular growth. The C50T9 case shows clear error growth for large $\eta$, but the central density evolution is unchanged. The slow collapse cases have larger energy error magnitudes and clear secular growth, corresponding to their significantly $\eta$-dependent central density evolution. As a rough estimate, the central density appears to track the converged evolution as long as the energy error magnitude stays below 1\%. It is possible that the energy error is proportional to the number of timesteps, explaining why long simulations experience greater error than ones on cosmological timescales.

The correlation between $\eta$ dependence, energy error growth, and collapse time suggests a connection between these characteristics. While a detailed analysis of this connection will be discussed in future work, the results shown here indicate that secular energy error growth can be a tracer for this particular numerical effect. These data also suggest that simulations with long collapse times are more susceptible to this accelerated collapse. Despite their numerical issues, long collapse times are important. Their performance can help diagnose the origins of numerical error that may play a role in shorter simulations, making them important for N-body algorithm building and testing. While our results so far show that the timestep size matters less for cosmological collapse than longer collapse times, further tests of short collapse times are needed to confirm our results are converged. These tests include varying the scattering probability limit by altering \texttt{DTimeFac} (Equation \ref{eqn:dtsidm}), and the use of other timestep definitions.

\section{Summary and Discussion}

\label{sec:conclusion}

In this work, we perform numerical tests of self-interacting dark matter halos using the \texttt{Arepo} N-body code. These tests study the influence of several numerical parameters and random initial condition noise on the core-collapse process. The effect of realization noise is studied by selecting reasonable $r_{\rm soft}$ and $\eta$ values based on preliminary convergence investigations, and generating 10 initial conditions for each resolution, with each realization using a different random seed. In addition to these realization noise studies, we create a grid of numerical parameter values, varying particle count, softening length, and timestep size, and study the effect that each combination of values had on the halo's central density evolution. This process is repeated for four cases, consisting of combinations of two initial halo concentrations and two core-collapse times. Different collapse times for halos with identical concentrations are generated by changing the scattering cross-section. From the results of these convergence tests, we find that our parameter selection in the realization noise investigation was reasonable.

Based on these studies, we identify several numerical effects that can be mitigated through informed parameter choices.
\begin{itemize}
    \item \emph{Resolution and realization noise:} When generating random initial conditions using \texttt{SpherIC}, halos with particle counts of $10^4$ suffer from significant scatter in evolution times (Figures \ref{fig:noise_C50T9} and \ref{fig:noise_C10T225}), as well as occasional outlier realizations that collapse much more slowly than the typical case. These outlier cases in our study are correlated with initial halo energies that are low in magnitude, suggesting a sensitivity to halo energy at low resolutions. The realization noise problem persists until a resolution of $10^5$ particles, where most simulations appear to converge and have significantly reduced scatter in core collapse time (Figure \ref{fig:noise_tcplot}). Our C50T225 case (high concentration, slow collapse) has not converged at this resolution, though, prompting caution when simulating similar halos. The $N=10^5$ requirement is of particular importance for simulations including substructure, which may include subhalos that are not sufficiently well-resolved to accurately resolve gravothermal core-collapse.
    \item \emph{Gravitational softening:} We find that Equation \ref{eqn:soft}, the CDM softening length presented by \cite{vdB2018}, is an appropriate softening choice in all cases tested. Selecting softening lengths larger than given in Equation \ref{eqn:soft} leads to a delayed collapse process (Figure \ref{fig:resolutionTest}). The magnitude of this collapse delay depends on the resolution, with higher resolution simulations being more resilient to the effect. Lowering the Equation \ref{eqn:soft} softening value by a factor of $10^{-0.4}$ does not substantially alter the halo evolution, suggesting that oversoftening is more of a concern than undersoftening. While this prescription for gravitational softening works well for our isolated halo, it is not immediately clear how to proceed in cosmological simulations that include many halos of varying mass and concentration. We expect that each individual halo would perform well with softening assigned by Equation \ref{eqn:soft}, making a global softening treatment difficult. Additionally, simulations including substructure present additional complications in softening length selection, as subhalos lose mass over time due to tidal effects.
    \item \emph{Timestep size:} Simulations that run far longer than the age of the Universe are sensitive to the $\eta$ parameter, which scales the simulation's gravitational acceleration-based timestep. Using the timestep definition in Equation \ref{eqn:dta}, a value of $\eta=0.025$ is sufficient for short simulations, but when simulating beyond the age of the Universe this value leads to energy non-conservation and an altered central density evolution (Figure \ref{fig:eta_evolution}). This error typically manifests as an accelerated core-collapse, but the C50T225 simulation is once again an outlier, with one high resolution simulation of that halo instead suffering from a significantly delayed collapse. The effects of a large $\eta$ value are also more severe for lower resolutions, adding another dimension to parameter selection. The cause of this error is not yet understood, but will be studied in future work. While the cause is not clear, the change in the total system energy consistently traces the issue and can be used as a diagnostic. We find that selecting $\eta=0.005$ significantly reduces the error in our long simulations compared to using $\eta=0.025$, but may not be fully converged, even with $N=10^6$ particles.
\end{itemize}

Throughout our convergence testing, the C50T225 case is a consistent outlier that struggles to converge with every numerical parameter. With its high concentration $c_{200m}= 50$ and lengthy 225 Gyr collapse time, this case does not represent a typical halo in realistic cosmological simulations. While this halo is an extreme case, a more subtle instance of these errors may be present in more common regions of the parameter space, making a thorough understanding of \texttt{Arepo}'s limits important.

The reason for the C50T225 non-convergence is not clear, but there are some possibilities. This case has the most significant difference between the characteristic timescales for gravity and scattering; the high concentration leads to large accelerations and short gravitational timesteps in the center of the halo, while the low cross-section causes large intervals of time between scattering events. Since \texttt{Arepo} adopts the smaller of the acceleration timestep and SIDM timestep, this leads to timesteps with very low scattering probabilities. This difference can be seen in the small value of $\hat{\sigma}=0.00827$, which tells us that this halo is deep in the lmfp regime, where gravitational interactions vastly outweigh scattering events. Hypothetically this regime should be straightforward to simulate as it is the closest case to CDM, but these results suggest a numerical issue in the \texttt{Arepo} scattering algorithm when the gravitational and scatterng characteristic times are very different. We plan a more detailed investigation of this error in future work, where we investigate the cause and determine if it is specific to \texttt{Arepo}.

The other recent works shown in Figure \ref{fig:CT_regimes} share some of our findings. Palubksi \emph{et al.} discovered similar energy non-conservation as our study, and found that selecting a stricter collision probability limit (\texttt{DTimeFac}) and a small timestep parameter ($\eta$) alleviates the issue~\cite{palubski2024numerical}. Fischer \emph{et al.} also discovered an energy error, and finds that switching to a fixed timestep value and using a smaller gravitational tree opening angle (\texttt{ErrTolForceAcc}) reduces the issue~\cite{fischer2024numerical}. The success of fixed timestepping in Fischer \emph{et al.} suggests that the energy error may indeed be caused by the non-symplectic nature of an N-body simulation with scattering events, which was one of the concerns motivating this work.

By quantifying the above numerical effects in SIDM N-body simulations using \texttt{Arepo}, future studies can identify and avoid similar complications. More important than any specific guidance on parameter selection given here is the demonstration that the introduction of self-interactions into N-body simulations changes the behavior of numerical errors, and that parameter choices suitable for CDM may no longer provide converged and useful results. Further work is needed to fully understand how to select numerical parameters in SIDM N-body simulations. In particular, the causes and behavior of the $\eta$ dependence discovered in this work must be fully understood so that we can confidently say we have converged results for long simulations. The extent of these effects in halo substructure must also be explored, so that we can assess the accuracy of cosmological simulations that suggest SIDM as a solution to the halo diversity problem. Lastly, the studies done in this work would benefit from a detailed comparison of \texttt{Arepo} and other N-body codes, to determine to what extent these issues are code-dependent. Until there is a more detailed understanding of these errors and their causes, N-body SIDM simulators must systematically check that their choices of parameters are sufficient to produce the desired level of convergence for each specific application.


\section*{Acknowledgments}

We thank Akaxia Cruz, Tom Quinn, Igor Palubski, Manoj Kaplinghat, and Moritz Fischer for useful discussions and feedback.

This work was supported in part by the NASA Astrophysics
Theory Program, under grant 80NSSC18K1014, and by a grant from the OSU College of Arts and Sciences Division of Natural and Mathematical Sciences. Z.C. Zeng is partially supported by the Presidential Fellowship of the Ohio State University Graduate School. The simulations in this work were conducted on Ohio Supercomputer Center \cite{osc}, mostly on the CCAPP condo. This research was supported in part by grant NSF PHY-2309135 to the Kavli Institute for Theoretical Physics (KITP)

\section*{Data Availability}

The N-body simulation snapshots generated in this work are available in the Dryad online data repository \cite{dryadData}. This public datasets includes central density evolution data, as well as N-body particle data. To meet the storage constraints of the repository we omit some snapshots of the $N=10^6$ simulations, and all of the realization noise simulations in section \ref{sec:noiseTests}. More complete data may be available on request to the authors.

\appendix

\section{C10T9 and C50T225 Convergence Tests}

\begin{figure*}
	\includegraphics[width=2\columnwidth]{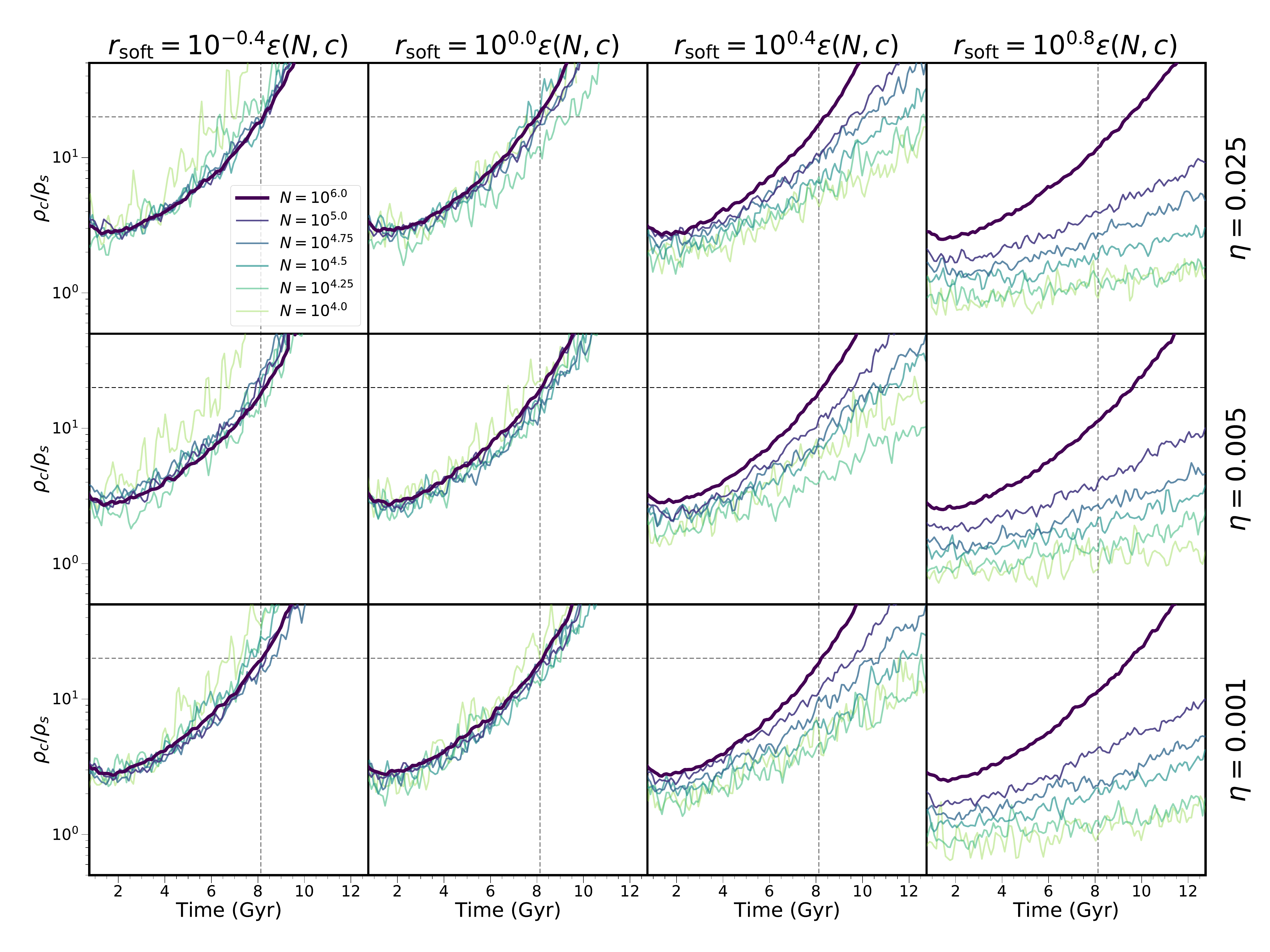}
    \caption{Convergence testing results for halo C10T9. Each panel shows the central density evolution at various resolutions. Panels in the same column share the same softening parameter, and panels in the same row share the same $\eta$, as labeled on the top and right sides of the figure respectively. The $N=10^6$ simulation line is bold in each panel. The horizontal dashed lines mark $\rho_c/\rho_s=20$, our core-collapse threshold, and the vertical dashed lines mark the core-collapse time of our $r_{\mathrm{soft}}=\epsilon(N,c)$, $\eta=0.001$, $N=10^6$ case.}   \label{fig:gridC10T9}
\end{figure*}

\begin{figure*}
	\includegraphics[width=2\columnwidth]{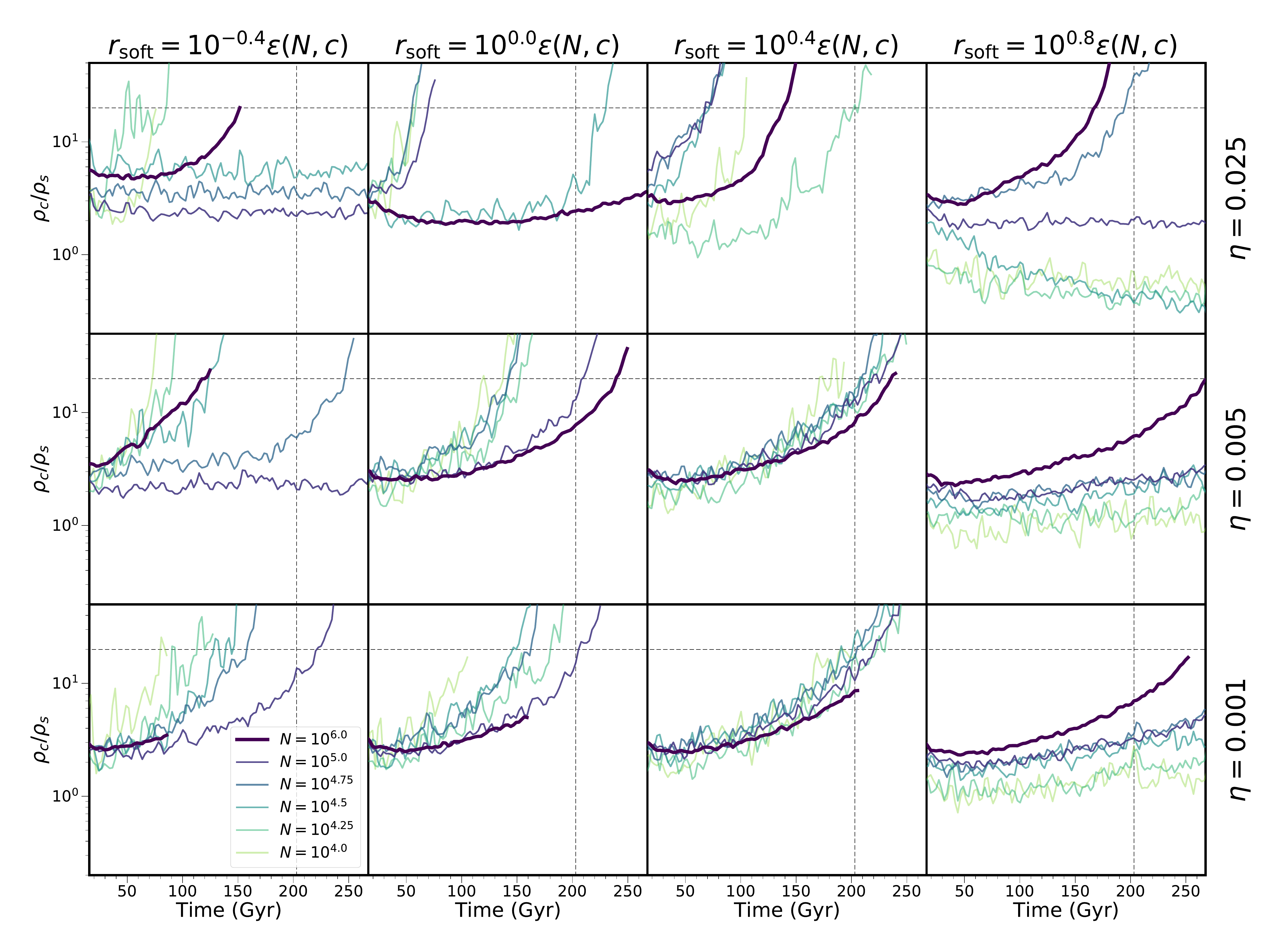}
    \caption{Convergence testing results for halo C50T225. Each panel shows the central density evolution at various resolutions. Panels in the same column share the same softening parameter, and panels in the same row share the same $\eta$, as labeled on the top and right sides of the figure respectively. The $N=10^6$ simulation line is bold in each panel. The horizontal dashed lines mark $\rho_c/\rho_s=20$, our core-collapse threshold, and the vertical dashed lines mark the core-collapse time of our $r_{\mathrm{soft}}=\epsilon(N,c)$, $\eta=0.001$, $N=10^5$ case.}   \label{fig:gridC50T225}
\end{figure*}

The C10T9 and C50T225 equivalents of Figures \ref{fig:gridC50T9} and \ref{fig:gridC10T225} were omitted from the main text for brevity. Here they are included to show to full range of simulations completed as part of this work.

Figure \ref{fig:gridC10T9} shows the convergence tests for the C10T9 halo. This figure shows the same trends as Figure \ref{fig:gridC50T9}, including the lack of $\eta$ dependence at any softening length.

Figure \ref{fig:gridC50T225} shows the convergence tests for the C50T225 halo. Due the higher computational cost of this halo at high resolutions and low $\eta$ values, the $N=10^5$, $\eta=0.001$, $r_{\mathrm{soft}}=10^0\epsilon(N,c)$ case is used instead of $N=10^6$ for the collapse time reference line. Some expected trends are visible in this figure, such as the delayed collapse at large softening lengths. We also see a general trend of collapse acceleration at low $\eta$ values, which is more significant for low resolutions. There are, however, significant outliers to these trends. As discussed in the context of Figure \ref{fig:eta_evolution}, the $N=10^6$ simulation for $\eta=0.025$ and $r_{\mathrm{soft}}=10^{0}\epsilon(N,c)$ is not accelerated as are most of the other large $\eta$ simulations, but instead significantly delayed. We also see that when large $\eta$ is combined with oversoftening ($\eta=0.025$ and $r_{\mathrm{soft}}=10^{0.8}\epsilon(N,c)$) low resolution data not only suffers delayed collapse, but forms very low density cores that are not expected from the gravothermal fluid model. Another unusual effect apparent in this figure is seen in the undersoftened simulations ($r_{\mathrm{soft}}=10^{-0.4}\epsilon(N,c)$), which show delayed core-collapse at low resolutions and large $\eta$. These unexplained trends may show the interplay of multiple numerical errors, causing unexpected behavior and non-convergence.
\bibliography{apssamp}

\end{document}